\documentclass[paper,notoc]{JHEP} 
\usepackage{epsfig}
\usepackage{amsmath}
\usepackage{amssymb}
\usepackage{cite}
\psfull

\def\kk{{\bf k_1}}
\def\kg{{\bf k_2}}

\def\pom{{I\!\!P}}

\title{Truncated BFKL Series in Electron-Proton Collisions}
\author{M.B. Gay Ducati and M.V.T. Machado\\  
  Instituto de F\'{\i}sica, Univ. Federal do Rio Grande do Sul.
 Caixa Postal 15051, 91501-970 Porto Alegre, RS, BRAZIL. \\
  E-mail: \email{gay@if.ufrgs.br}, \email{magnus@if.ufrgs.br}}

\abstract{ We study the contribution of the truncated BFKL Pomeron series to
the  electroproduction process  showing that a reliable description is
obtained taking into account two orders in perturbation theory. Using the
recent $F_2$ logarithmic slope data as a constraint to the adjustable
parameters, the inclusive structure function is described in a wide range of
the small $x$ HERA kinematical region, consistent with the unitarity bound. We
also extrapolate the predictions to the THERA region.}

\preprint{ }

\begin{document}
\section{Introduction}

Deep inelastic electron-proton scattering experiments at HERA have provided
measurements of the inclusive structure function $F_2(x,Q^2)$ in very
small Bjorken variable $x$ ($10^{-2}$ down to $10^{-5}$) \cite{Osaka2000}. In
these processes the proton target is analysed by a hard probe with virtuality
$Q^2=-q^2$, where $x \sim Q^2/2p.q$ and $p,\,q$ are the four-momenta of the
incoming proton and the virtual photon probe. In that kinematical region, the
gluon is the leading parton driving the small $x$ behavior of the deep
inelastic observables. The small $x$ region is described in a formal way using
the summation of gluon ladder diagrams, whose virtual contributions lead to the
gluon reggeization \cite{reggeization}. Such ladders diagrams are associated
with the Pomeron, a Reggeon  with the vacuum quantum numbers,  which
was introduced phenomenologically to describe the high energy behavior of
the total and elastic cross-sections of the hadronic reactions \cite{predazzi}
 and connected with the existence of large rapidity gaps in the produced final
state \cite{Bjorken}.

In leading order (LO) all powers of $\alpha_s\ln(Q^2/\mu^2)$, with $\mu^2$
the factorization scale, are summed by the DGLAP evolution equations
\cite{DGLAP}, which take into account only the  strongly ordered parton
transverse momenta ($k_T$) ladders. At present the next-to-leading order (NLO)
contribution is also used, including non-ordered $k_T$ contributions in a
covariant way \cite{GLAPNLO}. In the current HERA kinematical regime the DGLAP
approach is quite successfull, although the theoretical expectation of
deviations due to the  parton saturation phenomena
should be present \cite{Mueller}. Further, corrections leading to non-linear
evolution equation of the gluon distribution have been calculated in the
literature \cite{GLR,AGL}.

Moreover, at very small $x$ the leading logarithms $\alpha_s\ln(1/x)$ are
shown to be  important \cite{bfkl}. In the leading logarithmic
approximation (LLA) the QCD Pomeron corresponds to the sum of ladder diagrams
with reggeized gluons along the chain, which are strongly ordered in momentum
fraction $x$. Such sum is described by the Balitzkij-Fadin-Kuraev-Lipatov
(BFKL) equation \cite{bfkl}. The corrections at next-to-leading level (NLLA)
are now known, leading to strong modifications in the LLA spectrum \cite{NLO}.
Although some existing reasonable  direct comparison with data \cite{bfkdata},
the BFKL approach presents the diffusion phenomena on momentun transfer ($\ln 
k_T$), which turns out the description of the inclusive observables as the
structure function $F_2$ unreliable \cite{diffusion}. This comes from the fact
that the diffusion leads to an increasingly large contribution from the
infrared and ultraviolet regions of the transverse momenta $k_T$ where the
BFKL is not expected to be valid. 

An important characteristic present at both DGLAP and BFKL approaches is that
their solutions grow like a power of the center of mass energy $s$, therefore
violating the unitarity bound at very high energies. This is one of the major
problems of small $x$ physics, since the Froissart-Martin bound states 
that the total cross section should not rise faster than $\sim \ln^2(s)$ at
asymptotic high energies \cite{unitarity}. In the DGLAP approach, the
unitarity problem requires corrections to give rise to the saturation
\cite{Golec,Victor} (taming the growth of the gluon distribution), while in the
BFKL approach this difficulty can be surpassed  considering the  resum of all
multiple BFKL Pomeron (calculated at LO) exchanges in the total cross section
(and structure function) \cite{kovchegov}.

In this scenario, since the BFKL approach is asymptotic
itself we have proposed that a truncated BFKL series was used to describe both
total and differential cross section in hadronic collisions, namely
proton-(anti)proton reactions \cite{trunk1}. In the atainable  energies there
is no room on pseudorapidity to enable a completely resummed n-rung  ladder
and studies have reported  a strong  convergence of the BFKL series
considering few orders in the expansion \cite{ryskin}. Furthermore, it was
found in \cite{ryskin} important evidence that the asymptotic solution to the
BFKL equation is inappropriated in the most of the HERA range, and the
expansion order by order allows to identify the onset of the region where the
full BFKL series resummation is required. 

The resulting scattering
amplitude in Ref.  \cite{trunk1} is consistent with the unitarity bound and
obtained by just two orders on perturbation theory.  The first term in the
expansion (Born level) corresponds to the two reggeized gluon exchange leading
to a constant ($s$-independent) contribution to the total cross section. The
second term comes from the one rung gluon ladder and provides a logarithmic
enhancement to the cross section. These two contributions are enough to
describe successfully  the acellerator data, and  even the cosmic ray
measurements. The distribution in momentum transfer $t$, namely the elastic
differential cross section, is obtained taking into account the non-forward
amplitude. It was shown in \cite{trunk1} the importance of a reliable
phenomenological input for the proton impact factor, i.e., the coupling of the
ladder to the proton vertex. A remarkable result in \cite{trunk1} is that the
factorizable character in energy $s$ and momentum transfer $t$ presents in the
non-forward amplitude allows an overall parametrization for the slope
$B(s)$. This result is independent of the details about the Pomeron-proton
coupling and holds by requiring that the amplitude is finite at $t$ equal zero.
Further, a phenomenological analyzes to the  diffraction cone data of
$pp(\bar{p})$ scattering has been made considering those ideas \cite{kontros},
corroborating our calculations.   

Such good results in the hadronic sector motivate the  analyzes of reliability
at electron-proton collisions, focusing the small $x$ region. In this work we
sum two terms of the perturbative series in order to obtain the imaginary part
of the DIS amplitude, and through the Optical theorem  the inclusive structure
function $F_2(x,Q^2)$. Using the latest HERA data (H1 and preliminary ZEUS) on
$F_2$-logarithmic slope \cite{preldata} to constrain the remaining  parameters
at the small $x$ region, we perform a broadly description (no free
parameter) of the structure function in the kinematical range of
$10^{-5}<x<10^{-2}$ and $0.35<Q^2<150$ GeV$^2$. We analyse the resulting
predictions for the upcoming THERA project, confronting them with the
unitarity bound for this observable and compare the present approach to the
complete BFKL series, which we treat later on.

This paper is organized as follows. In the next section one introduces the
main formulae and the results for the inclusive structure function. In the
Sec.~(3), we calculate the observable $dF_2/d\ln Q^2$, and  perform  a fit
to the updated HERA data on $F_2$-slope, determining the values of the
existing  parameters. As a consequence, the inclusive structure function is
obtained  straightforwardly  and compared with the HERA small $x$ results. The
extrapolation into the THERA kinematical regime is performed in the Sec.~(4).
In the last section we draw our conclusions.

\section{Small-$x$ DIS in Perturbation Theory}

In the center of mass frame, the deep inelastic lepton-nucleon scattering
process is depicted as the scattering of the incoming lepton through a large
angle, which radiates a highly virtual photon $\gamma^*$. The photon then
scatters inelastically off the incoming nucleon (proton). The total cross
section for the process $\gamma^* \,p \rightarrow X$, with $X$ meaning all
possible final states, is obtained via Optical theorem  from the imaginary
part of the elastic $\gamma^*\,p \rightarrow  \gamma^*\,p$ amplitude (see Fig.
1). For high $\gamma^*\,p$ center of mass energies the BFKL approach is
indicated to compute this cross section \cite{forshaw}.

\begin{figure}
\centerline{\psfig{file=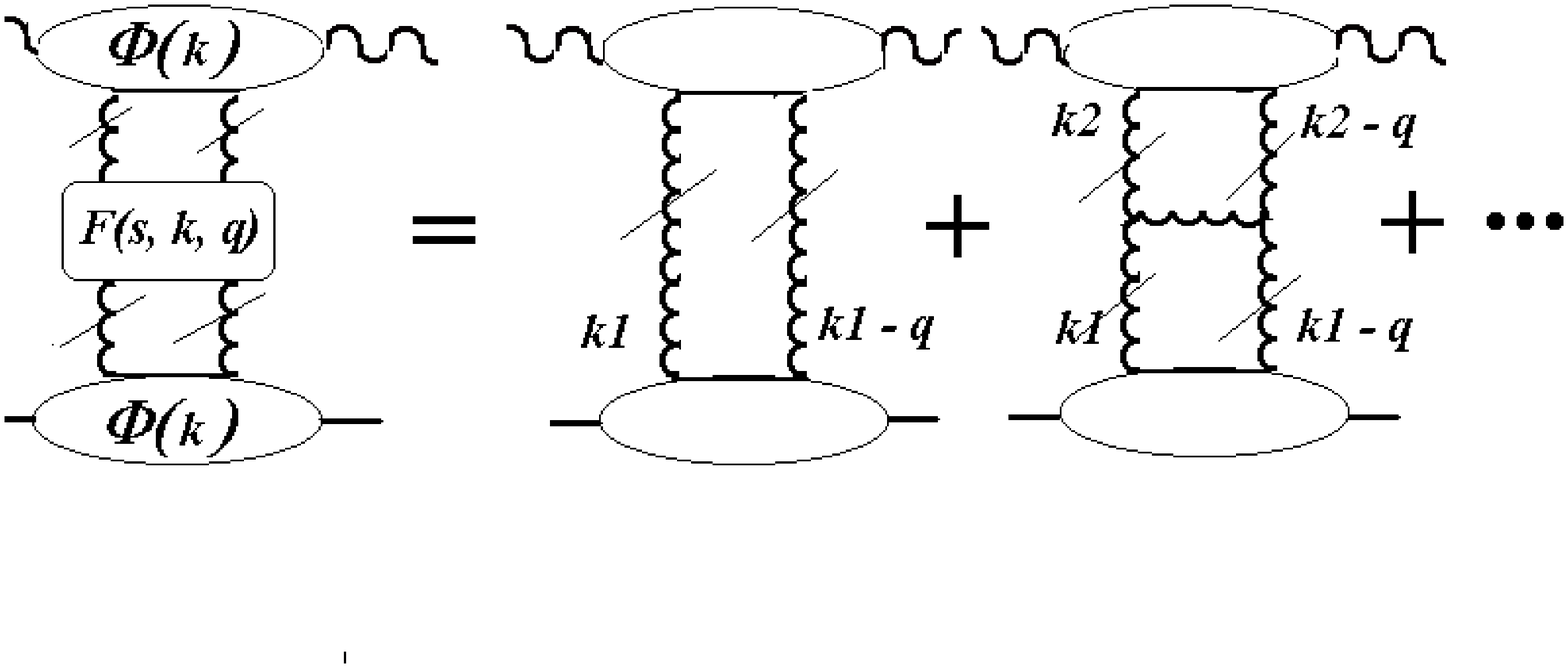,width=80mm}}
\caption{The upper blob denotes the virtual photon impact
factor, the lower one represents the proton impact factor and the first two
orders in perturbation theory are shown. In LLA, the ladder is constructed with
reggeized gluons in the $t$-channel and bare gluons on the $s$-channel (the
rungs), which are connected by a non-local gauge-invariant effective vertex
(the bold blob).} \end{figure}

Now let us define the kinematics for the DIS process. The incoming electron and
proton four-momenta are $l$ and $p$, respectively, and the virtual space-like
photon has a four-momentum $q$. The main kinematic invariants are given by

\begin{eqnarray}
Q^2=-q^2,\hspace{0.7cm} s=(p+l)^2,\hspace{0.7cm} W^2=(p+q)^2,\hspace{0.7cm}
x=\frac{Q^2}{2p.q} \approx  \frac{Q^2}{Q^2+ W^2}\,. \label{eq1}\end{eqnarray}

The last expression for $x$ is correct in the limit of negligible electron and
proton masses;  the high energy region $W^2 >>Q^2$, which implies $x<<1$,
defines  the small $x$ regime. The proton inclusive structure function and the
longitudinal structure function can be written in terms of the cross sections
for the scattering of transverse or longitudinal polarized photons in the form

\begin{eqnarray}
F_2(x,Q^2) &=& \frac{Q^2}{4\pi^2 \alpha} \left[\sigma_T(x,Q^2) +
\sigma_L(x,Q^2)\right] \,,\\
F_L(x,Q^2) &=& \frac{Q^2}{4\pi^2 \alpha}
\sigma_L(x,Q^2)\,. \label{eq2}
\end{eqnarray}

Then the structure functions are obtained computing the imaginary part of the
amplitude for elastic $\gamma^*\,p$ scattering considering each photon
polarization. In the asymptotic high energy limit, for photons with
polarization $\delta$, the cross section is the following \cite{forshaw}
\begin{eqnarray}
\sigma_{\delta}(x,Q^2)=\frac{{\cal G}}{(2\pi)^4} \, \int
\frac{d^2\kk}{\kk^2}\,\frac{d^2\kg}{\kg^2}\,\Phi^{\gamma^*}_{\delta}(\kk)\,
F(x,\kk,\kg)\,\Phi_p(\kg) \label{eq3}
\end{eqnarray}

Here, ${\cal G}$ is the color factor for the color
singlet exchange and $\kk$ and $\kg$ are the transverse momenta of the
exchanged gluons in the $t$-channel (see Fig. 1). The
$\Phi^{\gamma^*}_{\delta}(\kg)$ is the virtual photon impact factor (with
$\delta = T,\,L$) and $\Phi_p(\kk)$ is the proton impact factor. These
quantities correspond to the upper and lower blobs in the Fig (1),
respectively.

The impact factors $\Phi({\bf k})$, describing the interacting particles
transition in the particle-Reggeon (i.e, Reggeons meaning the reggeized gluons)
processes are, by definition,  factorized from the Green's function for the
Reggeon-Reggeon scattering $F(x,\kk,\kg)$.  As a consequence,
the energy dependence is determined by  this function, remembering that the
center of mass energy of the system $\gamma^*\,p$ is $W^2$ and that $x \approx
Q^2/W^2$.  The BFKL kernel $F(x,\kk,\kg)$  states the dynamics of the process
and is completely determined in perturbative QCD. For example, the cancelation of the infrared
singilarities in the kernel is known from Ref. \cite{FL}. In addition, the
amplitude describing the interaction of the particles (colorless) is the
convolution of the kernel with  the corresponding impact factors and it is
infrared safe. Moreover, the infrared singularities cancelation in the impact
factor of colorless particles has been demonstrated to next-to-leading order
in Ref. \cite{FM}.

The main properties of the LO kernel are well known
\cite{bfkl} and the results arising from the NLO calculations suggest that the
pQCD Pomeron can acquire very significant non-leading corrections, which  have
yielded intense debate in the current literature \cite{NLO}. The main
characteristic at LO is that the leading eingenvalue of the kernel leads to a
strong rise with decreasing $x$, $F(x)\sim \frac{x^{-\varepsilon}}{\sqrt{\ln
1/x}}$, where $\varepsilon=4\, \overline{\alpha}_s \ln 2 \approx 0.5$. As a
consequence, the inclusive structure function has this same growth at low $x$.
Therefore, the resulting amplitude (i.e. total cross section and structure
function) clearly violates the unitarity bound.

Instead of using the full BFKL series, we claim that a reliable description can
be obtained considering its truncation up to two orders in the perturbative
expansion. The arguments for this working hypothesis were already discussed in
the introduction but we recall them here. Such procedure has been
successfull into describing the hadronic cross sections in a consistent way
concerning the unitarity constraint \cite{trunk1} and may be considered in the
electroproduction case.  The justification is that in the current acellerator
domain the asymptotic regime was not reached and there is no room in rapidity
to enable an infinite n-gluon cascade, represented diagrammatically by the BFKL
ladder. Indeed, a steep convergence of the BFKL series in few orders in the
perturbative expansion  has been reported \cite{ryskin} and phenomenological
studies indicate that such a procedure is reasonable at least in proton-proton
collision \cite{fiore}.  In order to continue, we need to take into account
the convolution between the impact factors and the corresponding gluon ladder
exchanges in each order. The typical diagram for the DIS amplitude at small
$x$ in the lowest order in perturbation theory is depicted in Fig. (1). At the
leading order the amplitude at high energies is purely imaginary and written
at $t=0$ as

\begin{eqnarray}
{\cal A}^{Born}(W,t=0)=\frac{2\,\alpha_s \, W^2}{\pi^2}\,\sum_f e^2_f\, \int
\frac{d^2\kk}{\kk^4}\,\Phi^{\gamma^*}_{\bot} (\kk)\,\Phi_p(\kk)
\label{eq4}
\end{eqnarray}
and the next order in perturbation theory calculated at LLA yields
\begin{eqnarray}
{\cal A}^{NO}(W,t=0)= \frac{6\,\alpha^2_s \, W^2}{8\,\pi^4} \sum_f
e^2_f\,\ln(W^2/W^2_0)\, \int \frac{d^2\kk}{\kk^4}
\frac{d^2\kg}{\kg^4}\, \Phi_p(\kk)\,
K(\kk,\kg)\,\Phi^{\gamma^*}_{\bot} (\kg) \,, \label{eq5} \nonumber
\end{eqnarray} 

where the $\alpha_s$ is the strong coupling constant, which is
fixed in the LO BFKL approach.  The running of the coupling constant
contributes significantly for the NLO BFKL, since it is determined by
subleading one-loop corrections, namely the self-energy and vertex-correction
diagrams. The $W_0$ is the typical energy of the process, scaling the
logarithm of energy and take an arbitrary value in LLA. 

\begin{figure}[h]
\centerline{\psfig{file=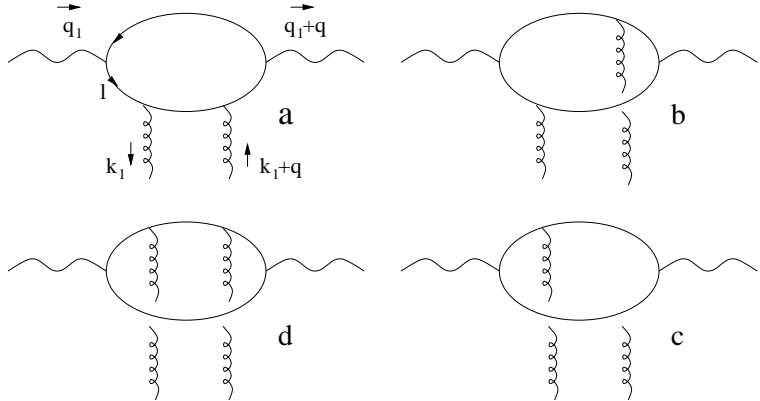,width=100mm}}
\caption{The diagrams contributing to the  virtual photon impact
factor at LO, from \cite{evanson}.} \label{diagsfot}
\end{figure}

The function $K(\kk,\kg)$ is the BFKL kernel at $t=0$  and describes the gluon
ladder evolution in the LLA of $\ln (s/t)$ as already discussed above.
The Pomeron is attached to the off-shell incoming photon through the quark
loop diagrams, where the Reggeized gluons are attached to the same and to
different quarks in the loop  \cite{evanson} (see Fig. \ref{diagsfot}).  The
$\Phi^{\gamma^*}_{\bot}$ is the virtual photon impact factor averaged over the
transverse polarizations \cite{balitsky}

\begin{eqnarray} \Phi^{\gamma^*}_{\bot}(\kg)=\frac{1}{2} \int_0^1
\frac{d\tau}{2\pi} \,\int_0^1 \frac{d \rho}{2\pi}\, \frac{\kg^2(1-2\tau
\tau^{\prime})(1-2\,\rho \rho^{\prime})}{\kg^2 \,\rho \rho^{\prime} + Q^2 \rho
\,\tau \tau^{\prime}} \,,
\label{eq6}
\end{eqnarray}
with $\rho$, $\tau$ the Sudakov variables associated
to the photon momenta and the notation $\tau^{\prime}=(1-\tau)$ and
$\rho^{\prime}=(1-\rho)$ is used. 

The diagrams at LO for the photon impact
factor are shown in the Fig. (\ref{diagsfot}). The average over longitudinal
polarization gives a smaller contribution to the total cross section (and
$F_2$), and by simplicity it will be  disregarded. 

Although the BFKL approach
is known at next-to-leading level \cite{NLO}, the impact factors are not yet
determined at this order of perturbatiove expansion.  Recently, it has been
reported the effect of the so called exact kinematics of the exchanged gluons,
which would give the most important contributions to the higher order terms
\cite{bialas}.

We are unable to compute the proton impact
factor $\Phi_p(\kk)$ using perturbation theory since it is determined by the
large-scale nucleon dynamics. A study has been  performed at Ref.
\cite{askew}, where the solutions of the Lipatov equation are critically
examined and their importance on the structure function
description is  determined using physically motivated modifications for small
${\bf k}^2$. Namely, it was performed a detailed parametrization of the
infrared region which satisfies the gauge invariance constraints when ${\bf
k}^2 \rightarrow 0$. Due to this gauge invariance $\Phi_p(\kk \rightarrow
0)\rightarrow 0$, and it can be modeled as a phenomenological input by the
expression \begin{eqnarray} \Phi_p(\kk)={\cal N}_p \,\frac{\kk}{\kk + \mu^2}
\label{eq7} \end{eqnarray}
where ${\cal N}_p$ is the unknown normalization of the proton impact factor
and $\mu^2$ is a scale which is typical of the non-perturbative dynamics.
Such  scale is related to the radius of the gluonic form factor of the proton. Considering
it as the scale of the hadronic electromagnetic form factor, then $\mu ^2
\simeq 0.5 $ GeV$^2$ instead of estimates from QCD sum rules giving $\mu ^2
\simeq 1-2$ GeV$^2$ \cite{askew}.

\begin{figure}[t]
\centerline{\psfig{file=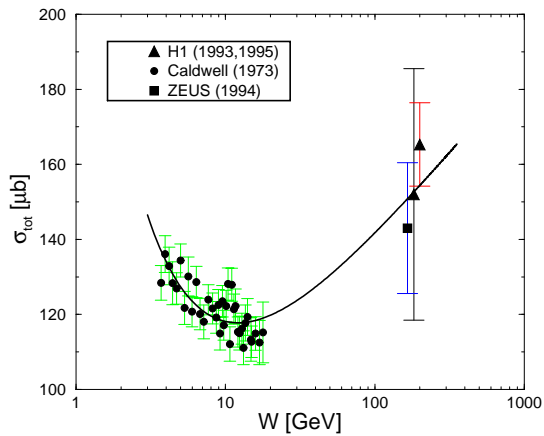,width=80mm}}
\caption{The result for the photoproduction cross section using the truncated
BFKL series compared with the data \cite{datafot}.} 
\label{phottot}
\end{figure}

When considering the photoproduction case there is no need
to deal with both a specific form for the impact factors and the transverse
momentum integration in the amplitudes. This allows to consider $W$-independent
factors in each term as free parameters and to obtain them from data on
photoproduction. In fitting these data, the Pomeron contribution described
above should be added through a small non-diffractive component arising from
the leading meson exchange trajectory. This is considered to be given by the
phenomenological Donnachie-Landshoff form \cite{DOLA} (first term in the
equation below). Then the expression for the photoproduction total cross
section is 

\begin{eqnarray} 
\sigma_{tot}^{\gamma^* \, p}= C_{R}\,(W^2/W^2_0)^{-0.4525} +
C_{Born} + C_{NO}\,\ln (W^2/W^2_0)\,. 
\end{eqnarray}
where $W^2_0=1$ GeV$^2$ (fixed), $C_R=0.216$ mb, $C_{Born}=0.044$ mb and
$C_{NO}=0.01$ mb. The result is shown in the Fig. (\ref{phottot}) with  the
cross section data \cite{datafot}.

In the electroproduction case,  using the Optical theorem and
summing the two orders in perturbation theory  we can write the expression for
the inclusive structure function, following \cite{balitsky}
\begin{eqnarray}
F_2(x,Q^2)= \frac{8}{3}\,\frac{\alpha^2_s}{\pi^2} \sum_f e^2_f \, {\cal N}_p
\left(I_{Born}(Q^2,\mu^2) + \frac{3\,\alpha_s}{\pi}\,\ln
\left(\frac{x_0}{x}\right) \,I_{NO}(Q^2,\mu^2)\right)\,,
\label{eq8}
\end{eqnarray}
where the functions $I_{Born,\,NO}(Q^2,\mu^2)$ are given by
\begin{eqnarray}
& & I_{Born}(Q^2,\mu^2)  = \frac{1}{2} \ln^2 \left( \frac{Q^2}{\mu^2} \right) +
\frac{7}{6}\ln \left( \frac{Q^2}{\mu^2} \right) + \frac{77}{18} \,,\\
& & I_{NO}(Q^2,\mu^2)  =  \frac{1}{6} \ln^3 \left( \frac{Q^2}{\mu^2} \right) +
\frac{7}{12}\ln^2 \left( \frac{Q^2}{\mu^2} \right) + \frac{77}{18}\ln \left( \frac{Q^2}{\mu^2} \right) +
\frac{131}{27} + 2\,\zeta (3)\,.\nonumber
\label{eq9} 
\end{eqnarray}
Here, the $x_0$ gives the scale to define the logarithms of energy for the
LLA BFKL approach and $\zeta (3)=\sum_r(1/r^3) \approx 1.202$ is the Riemann
$\zeta$-function. It was verified in \cite{balitsky} that the coefficients in
front of the logarithms of virtuality $Q^2$, which are determined by the
anomalous dimensions of twist-2 operators (OPE), coincide up to the factor
$2/3$. Also the subsequent two terms in the expansion, corresponding to the
contributions of order $\alpha_s^2 \ln ^2(1/x)$ and $\alpha_s^3 \ln ^3(1/x)$,
are calculated.

Some comments about the result above  are in order. The high energy or
Regge-limit for the virtual photon-proton scattering  is connected with the
behavior of the structure function $F_2(x,Q^2)$ when $x$ goes to very small
values ($x \rightarrow 0$). The main approaches to describe the Regge limit
are based on QCD perturbation theory \cite{DGLAP,bfkl}, and both  predict a
rise of the structure function at large $Q^2$ and small $x$. However, one
expects \cite{Mueller} this rise of the cross section at large energies to be
tamed by screenning corrections leading to an asymptotic behavior required by
the Froissart bound \cite{unitarity}. The result for $F_2$ in our calculations
is obtained taking into account this unitarization requirement {\it a
priori}.  

Concerning the Regge phenomenology, the truncated BFKL series presents the main
characteristics of the Regge-Dipole Pomeron model
(see, for instance \cite{dipolepom}). Namely, in the Dipole Pomeron
$F_2(x,Q^2)\sim R(Q^2)\ln (1/x)$ and this behavior  corresponds to the
contribution of a double $j$-pole to the partial amplitude of $\gamma^* p
\rightarrow \gamma^* p$, where $R(Q^2)$ is the residue function of the
Pomeron. The resultant trajectory has unit intercept, $\alpha_{\pom}(0)=1$.

Also it is interesting to compare the resultant structure function with the
complete perturbative expansion. Indeed, Navelet et al. \cite{navelet} have
taken into account the full BFKL series  described in the QCD dipole picture.
Using the $k_T$-factorization properties of  high energy hard interactions the
proton structure function is obtained in a 3-parameter expression, which gives
a fair description of H1 data at $Q^2 \leq 150$ GeV$^2$. It is written as
\begin{eqnarray} F_2^{BFKL}(x,Q^2)={\cal N}\, a^{1/2}
\exp^{(\alpha_{\pom}-1)\ln \left( \frac{1}{x} \right)}
\left(\frac{Q}{Q_0}\right) \exp^{-\frac{a}{2}\ln ^2 \left( \frac{Q}{Q_0}
\right)} \label{eq10} \end{eqnarray}

There, $a=\left(\frac{\bar{\alpha}_s N_c}{\pi} 7\,
\zeta(3)\ln(1/x)\right)^{-1}$, and $\alpha_{\pom}=4\,\bar{\alpha}_s N_c \ln
2/\pi + 1$ is the BFKL Pomeron intercept. The parameters are ${\cal N}=0.077$,
$Q_0=0.627$ and $\alpha_{\pom}=1.282$ (corresponding to an effective coupling
constant $\bar{\alpha}_s \simeq 0.11$).  Moreover, in Ref. 
\cite{vicgay} the DLA limit of BFKL in the dipole picture  was
calculated and it was shown that the transition region between these two
approaches is $Q^2 \approx 150$ GeV$^2$, i.e. this is the domain where the LO
BFKL Pomeron states. Later, we turn out to the Navelet et al. result in our
predictions to the THERA kinematical region, when one compares both the
truncated and the full BFKL series in that domain.

As a remark, we notice that  the  expression (\ref{eq8}), consistent with the
unitarity limit, gives some theoretical fundamentation to the existing study
regarding  scaling properties in deep inelastic scattering \cite{haidnt}.
It was found  that the available data on small $x$ regime are well described
by a polinomial linear in $\ln (1/x)$ for every measured $Q^2$-value, $
F_2(x,Q^2)=u_0(Q^2) + u_1(Q^2) \log \frac{\nu(Q^2)}{x}$,  where possible
higher order terms in $\log (1/x)$ are statistically non significant
\cite{haidnt}.

Now, having the expression for the inclusive structure function, in the next
section we compare it with the HERA experimental results, determining the
adjustable parameters and the range of validity from this model.

\section{Comparison with the HERA experiment}

Now we proceed to compare the expression obtained for the inclusive structure
function $F_2(x,Q^2)$ with the truncated BFKL series with the experimental
results. We choose to use a smaller data set defined by the latest HERA
experimental  measurements of the $Q^2$ logarithmic derivative of $F_2$, which
lies in the small $x$ region $x<10^{-2}$.  These updated data on the slope are
precise enough to determine unambiguously the two phenomenological
non-perturbative parameters ${\cal N}_p$, $\mu^2$ and the scale $x_0$.

Before to perform the analyzes, we should discuss the main features of the
slope data.  The global analyzes of the proton structure function data do not
present significant deviations from DGLAP approach, however measurements of
the rate of change of the logarithmic $Q^2$ dependence of $F_2$, i.e. the
$Q^2$-slope, have shown possible deviations from the expected pQCD behavior at
small $x$ and small $Q^2$. These results may be interpreted as signal of
unitarity corrections, namely that the system has reached to a QCD regime of
gluon saturation \cite{vicsat,levslope}. Such effects can be analyzed in that
quantity, since in leading order DGLAP evolution it is directly proportional
to the gluon structure function $xG(x,Q^2)$ \cite{pritz},
\begin{eqnarray}
\frac{\partial F_2(x,Q^2)}{\partial \ln
Q^2}=\frac{2\,\alpha_s}{9\,\pi}\,xG_{DGLAP}(x,Q^2)\,.
\label{gluon}
\end{eqnarray}

The most important experimental observation is that at fixed $x$ the slope is
a monotonic decreasing function of the virtuality $Q^2$. For fixed $Q^2$ it
increases as $x$ becomes smaller. In the reference \cite{levslope}, it was
verified that the $Q^2$-slope behavior as measured in the kinematic region
currently available at HERA does not present unambiguous signs of saturation.
Indeed, this observable can be described by the usual DGLAP formalism
considering the latest parton distribution or  DGLAP with saturation effects
as well as  by some Regge based models. The procedure to distinguish between
models for gluon saturation and the Regge inspired ones (including soft and
hard interactions) is  not an easy task because the saturation scale
$Q_s(x)\sim 1-5$ GeV$^2$ (setting the virtuality where unitarity effects start
to be visible) at the current HERA domain occours at similar values of the
matching of the soft and hard components. Indeed, concerning this question we
have proposed the logarithmic slope from the diffractive structure function as
a potential observable to distinguish such dynamics \cite{difder}.

Considering the truncated BFKL series, the $F_2$-slope can be calculated
straightforwardly from Eq. (\ref{eq8}), yielding
\begin{eqnarray}
\frac{\partial F_2(x,Q^2)}{\partial \ln
Q^2} & = & \frac{4}{3}\,\frac{\alpha^2_s}{\pi^2} \,\sum_f e^2_f {\cal N}_p\,
\left[ 2\,\ln \left( \frac{Q^2}{\mu^2} \right) + \frac{14}{6} \, + \right.
\nonumber \\  & & + \, \left.
\frac{3\,\alpha_s}{\pi}\,\ln \left(\frac{x_0}{x}\right) \left( \ln^2 \left(
\frac{Q^2}{\mu^2} \right) + \frac{7}{3}\ln \left( \frac{Q^2}{\mu^2} \right) +
\frac{77}{9} \right) \right] 
\label{derivada}
\end{eqnarray} 

The above theoretical result is compared with the current HERA data for the
logarithmic slope. A combination of the published H1 data and the
preliminary ZEUS ones \cite{preldata} is used. Both sets of experimental
data  allow to cover a kinematical range corresponding to virtualities of 
$0.3 < Q^2 < 40$ GeV$^2$ and Bjorken variable of $10^{-5}<x<10^{-2}$.

\begin{figure}[t]
\centerline{\psfig{file=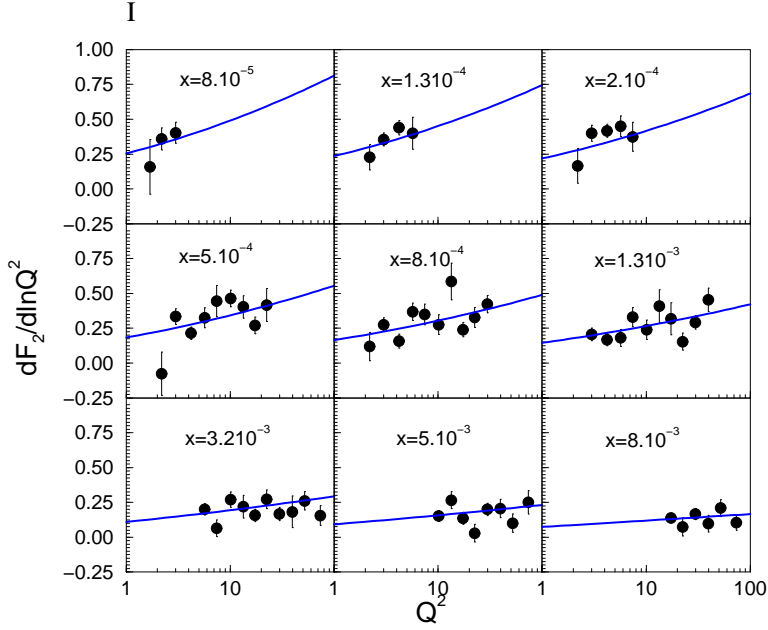,width=100mm}}
\caption{The result for the partial derivative $\partial F_2/ \partial
\ln Q^2$ plotted as a function of $Q^2$ (in GeV$^2$) for fixed $x$,  using the
truncated BFKL series compared with the latest H1 data \cite{preldata}. The
errors are summed in quadrature.}  \label{h1data} \end{figure}

In the Fig. (\ref{h1data}) we show the result for the H1 data
on the $Q^2$-logarithmic derivative, considering the $Q^2$ dependence  at fixed
$x$. A reasonable description is obtained for the measured kinematical range.
The resultant  parameters are shown in the Table 1. We noticed that to consider
the strong coupling constant as a free parameter does not improve the
description, therefore one considers it fixed, namely $\alpha_s=0.2$. The
overall normalization is defined as ${\cal
N}=\frac{4}{3}\,\frac{\alpha_s}{\pi^2} \,\sum e^2_f {\cal N}_p$, depending on
non-perturbative  physics through the normalization of the proton impact
factor ${\cal N}_p$.

\TABULAR{|c|c|c|c|c|}
{\hline DATA SET & ${\cal N}$ & $\mu^2$ & $x_0$ & $\alpha_s$ \\
\hline
H1 Collaboration & 0.009 & 0.068 & 0.008 & 0.2 (fixed) \\
\hline}{The parameters  from the H1 data on the
$Q^2$-logarithmic derivative \protect\cite{preldata} ($ \chi^2=0.01$).}


Although the quite different expression for the dependence on the virtuality
(see Eq. \ref{derivada}), our  approach provides a similar shape as the
parametrization linear in $\ln Q^2$ employed by the H1 Collaboration in the
data analyzes \cite{preldata}. The value found for the scale $\mu = 0.26$ GeV
is consistent with a non-perturbative input, i.e. $\mu \sim \Lambda_{QCD}
\approx 0.2$ GeV. The parameter $x_0$, which provides  the scale to define the
logarithms on energy and the region where the dynamics of the LLA
summation is suitable, lies in the small $x$ domain, $x \ll x_0 \approx
10^{-2}$.

\begin{figure}[t]
\centerline{\psfig{file=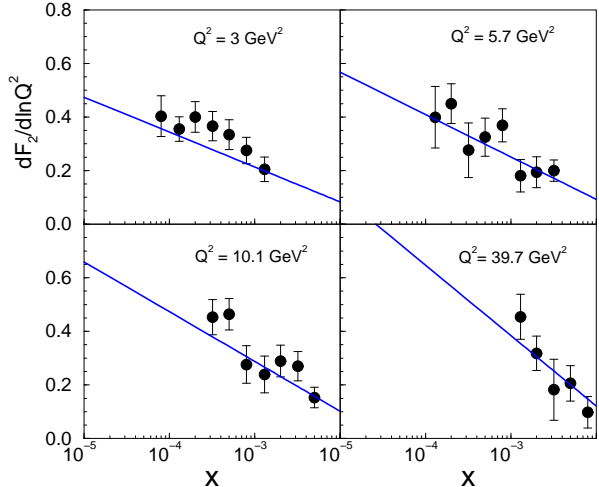,width=80mm}}
\caption{The result for the partial derivative $\partial F_2/ \partial
\ln Q^2$ plotted as a
 function of $x$ for fixed virtualities
$Q^2=3,\, 5.7,\, 10.1,\,39.7$ GeV$^2$,  using the truncated BFKL series
compared with the latest H1 data \cite{preldata}. The errors are summed in
quadrature.}
  \label{derq2} 
\end{figure}

The dependence on $x$ for the logarithmic slope is presented in Fig.
(\ref{derq2}), at fixed representative virtualities.
The model describes broadly the data in the range considered. An important
feature of the approach is a attenuated increasing of the gluon distribution at
low $x$ [see Eqs. (\ref{gluon}-\ref{derivada})], observed in the plots through
the extrapolations into smaller momentum fraction values. The DGLAP evolution
equation, instead provides a stronger and continuous rise towards small $x$ for
fixed $Q^2$ \cite{preldata}.  

\begin{figure}
\centerline{\psfig{file=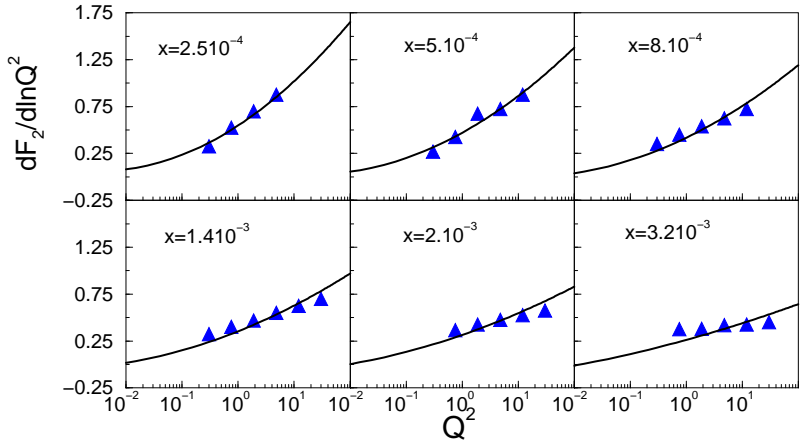,width=100mm}}
\caption{The result for the partial derivative $\partial F_2/ \partial
\ln Q^2$ plotted as a function of $Q^2$ (in GeV$^2$) for fixed $x$,  using the
truncated BFKL series compared with the latest preliminary ZEUS data. The
points were taken from the plots of \cite{preldata}.}  
\label{zeusdat}
\end{figure}

In the Fig. (\ref{zeusdat}) we show the result for the preliminary ZEUS data on
the logarithmic slope  versus $Q^2$, at  fixed momentum fraction $x$ . The
data  points were taken from  the plots of  \cite{preldata}. We determined
the adjustable parameters from this data set, however it should be stressed the
preliminary character of this result. The values are shown in Table (2).

\TABULAR{|c|c|c|c|c|}
{\hline DATA SET & ${\cal N}$ & $\mu^2$ & $x_0$ & $\alpha_s$ \\
\hline
ZEUS Coll. (preliminary) & 0.006 & 0.053 & 0.0053 & 0.2 (fixed) \\
\hline}{The parameters from the ZEUS preliminary data on the
$Q^2$-logarithmic derivative \protect\cite{preldata}.}


The values obtained are consistent with those found from the H1 data set.  The
main feature of the ZEUS experimental measurement is the presence of points
obtained at very low virtualities $Q^2 \geq 0.3$ GeV$^2$.  The plots present
the same characteristics as the H1 data set, for instance,  a linear behavior
in $\ln Q^2$  is observed. We notice that the ZEUS Collaboration parametrized
the slope data with the expression $\frac{\partial F_2}{\partial \log_{10}
Q^2}= B+ C\log_{10}Q^2$. In  our analyzes we converted the data points to
$\ln Q^2$,  similarly to  those taken from H1.

Having the parameters determined from the fit to the $Q^2$-slope data, we use
our expression for the inclusive structure function $F_2(x,Q^2)$ to determine
the region of validity of the model considering the truncation of the BFKL
series. The data set are separated, in the context of this paper, in the
regions corresponding to low, medium and large virtualities, considering both
H1 and ZEUS experimental results \cite{dataearlier}. Here, one uses the
parameters values obtained from H1 data set, since the analyzes are published.
The first set analysed was the low $Q^2$ data in HERA collider. This
kinematical domain has considerable importance since it is a region of
transition between soft and hard physics. Namely, it defines the range where
unitarity corrections should start to appear or the transition region for
the  matching of soft-hard contributions. The results are shown in the Figs.
(7-8).   \newpage
\begin{figure}[t] \centerline{\psfig{file=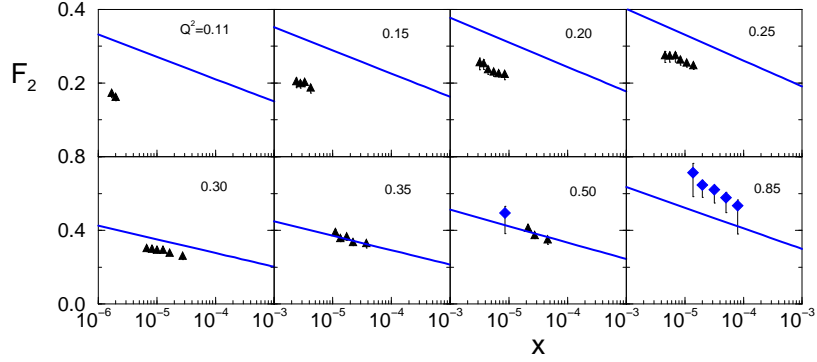,width=100mm}}
\caption{The estimates for the inclusive structure function at very small
virtualities (in GeV$^2$) with the data of H1 (diamonds) and ZEUS (triangles
up) \cite{dataearlier}.} \label{low1}
\end{figure}  \vspace{-2cm}
\begin{figure}
\centerline{\psfig{file=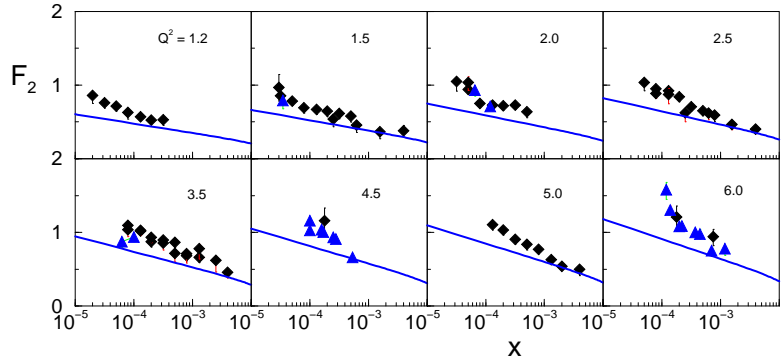,width=100mm} }
\caption{The estimates for the inclusive structure function at small
virtualities (in GeV$^2$) with the data of H1 (diamonds) and ZEUS (triangles up)
\cite{dataearlier}.} \label{low2}
\end{figure}

We notice that this is a prediction with no additional free
parameter, all already determined from the slope data [see Tab. (1)]. A
reasonable description is present starting at virtualities $Q^2>0.3$ GeV$^2$.
For $Q^2\geq 0.85$ GeV$^2$ the curve lies slightly below the data points. This
fact is expected, since even that the  main contribution at small
$x$ should come from the perturbative Pomeron, a certain amount of background
in the structure function due to quarks and non-perturbative effects (soft
exchanges like the Donnachie-Landshoff Pomeron) should be taken into
account \cite{bfkdata}. It is noticeable also the partial agreement with the
very low $Q^2$ measurements, where we would not expect such a  good
description.

The next set of data corresponds to the medium virtualities for the H1 and ZEUS
measurements \cite{dataearlier}. The result is shown in the Fig. (9-10). The
truncated BFKL series describes reasonably the data in this
kinematical range, except for excedingly larger $x$ values ($x>10^{-2}$). The
curve still lies slightly below the data points, but with better accordance
than the previous set. Deviations at greater $x$ are expected, since we
constraint our parameters from data belonging to $x<10^{-2}$ and the reggeonic
contribution should not be disregarded in this kinematical region. 

\newpage
\begin{figure}[h] 
\centerline{\psfig{file=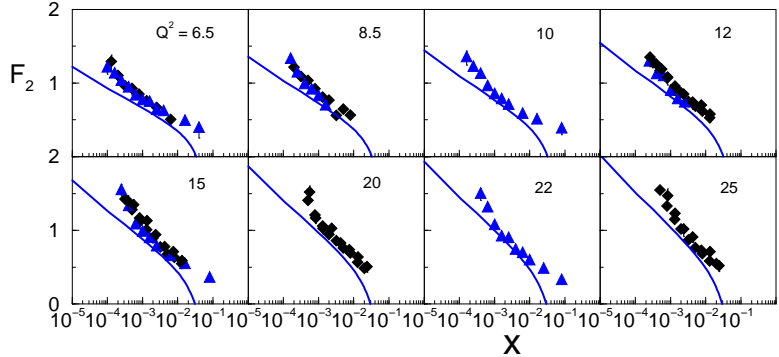,width=100mm}}
\caption{The result for the inclusive structure function at medium 
virtualities (in GeV$^2$) with the data of H1 (diamonds) and ZEUS (triangles
up).} \end{figure} 
\vspace{-2cm}
\begin{figure}[h] 
\centerline{\psfig{file=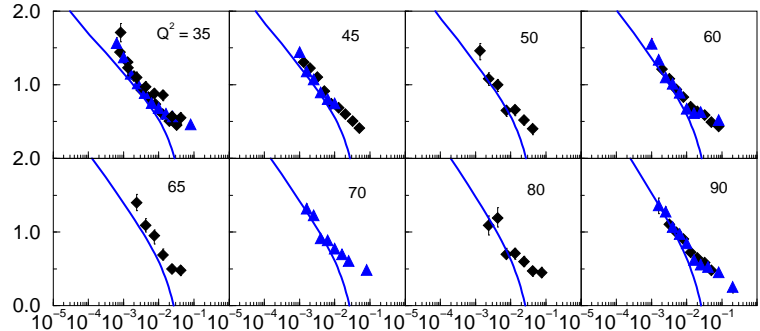,width=100mm}}
\caption{The result for the inclusive structure function at medium to high 
virtualities (in GeV$^2$) with the data of H1 (diamonds) and ZEUS (triangles up).} \end{figure} 

The last data set corresponds to the large virtualities range at HERA. The
results are shown in the Fig. (11). The truncated BFKL series reproduces the
data on $x<10^{-2}$ quite well up to $Q^2=350$ GeV$^2$. The data set on large
virtualities are dominated by data at large $x>10^{-2}$, which we are unable to
describe since this region is not dominated by the perturbative Pomeron. An
improvement would be the introduction of  the reggeonic piece, as discussed
above.

\begin{figure}[t] 
\centerline{\psfig{file=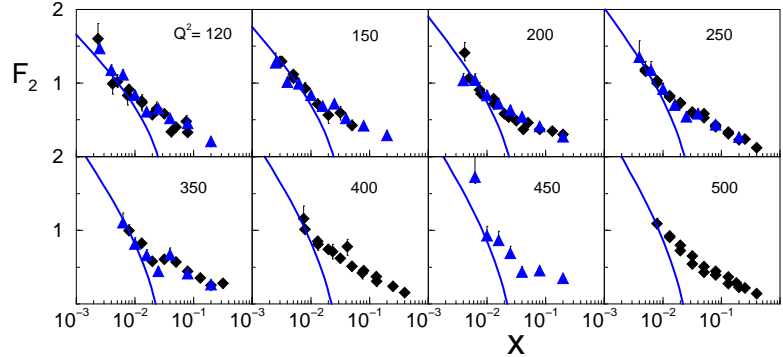,width=100mm}}
\caption{The result for the inclusive structure function at large 
virtualities (in GeV$^2$) with the data of H1 (diamonds) and ZEUS (triangles
up).} \end{figure} 

Summarizing the results presented for the proton structure function: 
\begin{itemize}
\item  the
truncated BFKL series up to second term allows to describe simultaneously the
slope data and the $F_2$ in a wide kinematical regime. The description of the
slope is in good agreement with the experimental results. Regarding the
structure function $F_2$, agreement is obtained for the range $0.3
\leq Q^2 \leq 350$ GeV$^2$, considering the kinematical constraint  $x \leq
10^{-2}$. If a such cut is not considered, the description is restricted in a
smaller range of virtualities ($0.3 \leq Q^2 \leq 150$ GeV$^2$);

\item  the growth of $F_2(x,Q^2)$ presents large
deviations from the steep increasing present at both the LO BFKL series and
the DGLAP approach, where $F_2 \sim x^{-\lambda}$;

\item there is room to include the contribution  from non-perturbative sector
(soft dynamics), mainly at low $Q^2$ virtualities. However it should be
stressed that they are not big, namely we estimated that the correction
corresponds to about 20 \% to the overall normalization; 

\item the measurements at very large $Q^2$ lie often at
$x > 10^{-2}$ where the approach is expected not to be valid and a
reliable parametrization to the large $x$ region should be introduced; 

\item we are not able to
distinguish unambiguously the complete BFKL series \cite{navelet} and the
truncation up to second order, since both approaches describe the current HERA
data. For instance, Navelet et al. and the present study produce a broadly
description of data at $1.5 \leq Q^2 \leq 150$ GeV$^2$. However, our analyzes
extend the predictions to very low $Q^2 \sim 0.3$ GeV$^2$ and larger $Q^2 \geq
150$ GeV$^2$.

\item  a study of the sensitivity of summing an additional term in the
perturbative expansion is required, and at this level the result should 
reach  the asymptotic behavior $F_2 \sim \ln^2 (1/x)$.

\end{itemize}
In the next section one uses the main results and conclusions obtained here, to
make predictions to the forthcoming experiment THERA. We focus in the behavior
of the complete BFKL series and its truncation in order to estimate the
$Q^2$ logarithmic slope and the proton structure function in that kinematical
domain.   

\section{The THERA kinematical Region}

The upcoming THERA project \cite{THERA}, that will be realized at DESY as the
successor of  HERA, should use the  electrons or positrons accelerated in
the linear collider TESLA and the protons from HERA. THERA enables the energy
frontier in deep inelastic physics to be raised up to about $\sqrt{s} \simeq
1$ GeV$^2$, allowing studies at virtualities $Q^2$ about $10^{6}$ GeV$^2$ and
Bjorken $x$ down to $10^{-6}$ in the perturbative region. This future
kinematical domain enables effectivelly to probe the saturation
phenomena, since the saturation scale at THERA would be higher than the values
found at HERA. 

\begin{figure}[h] 
\centerline{\psfig{file=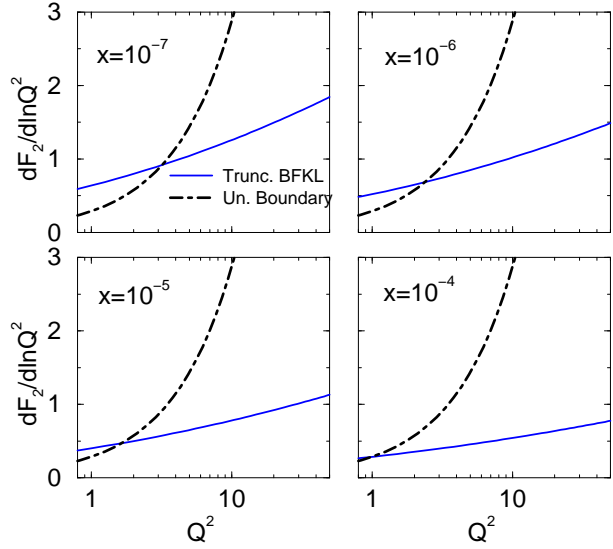,width=90mm}}
\caption{The prediction of the truncated BFKL series (solid lines) for the
$F_2$-slope versus $Q^2$ compared with the unitarity limit (dot-dashed lines)
in the THERA kinematical region.} 
\label{unbound}
\end{figure}

Our first study in the THERA kinematical region is to consider the 
logarithmic derivative, Eq. (\ref{derivada}), versus the virtualities $Q^2$,
considering the values of $x$ planned to be available (see Fig. \ref{unbound}).
This quantity can be compared with the unitarity bound for the $Q^2$-slope, 
which using the weak form of the unitarity constraint is expressed as \cite{ayalaPLB}
 \begin{eqnarray}
\frac{\partial F_2(x,Q^2)}{\partial \ln Q^2} \leq \frac{Q^2R^2}{3\,\pi^2} \,,
\end{eqnarray}
where the size of the target is $R^2=8.5$ GeV$^{-2}$ \cite{levthera}. 

We verify that our model does not violate strongly the unitarity bound,
overestimating that limit at $Q^2 \simeq 2.5$ GeV$^2$ for $x=10^{-6}$ and $Q^2
\simeq 3$ GeV$^2$ for $x=10^{-7}$. Concerning the gluon saturation approaches,
the intersection point between the usual DGLAP dynamics and the unitarity
bound defines the saturation scale $Q_s(x)$. At HERA regime this scale lies in
the same order of the models mixing soft and hard contributions. However, at
THERA domain we  expect to distinguish between these approaches
\cite{levthera}. Instead, in our case the violation of the unitarity bound
informs us that at THERA regime there is enough energy to enable the emission
of one more $s$-channel gluon in the BFKL ladder (two rungs) and one should 
include the next term of the perturbative expansion, resulting in  $F_2 \sim
\ln^2(1/x)$. We notice that such a contribution actually saturates the
Froissart bound and the expansion should stop at this level. In other words,
the virtualities in which the unitarity bound is violated provide the threshold
region where the present approach is applicable.  

\begin{figure}[t] 
\centerline{\psfig{file=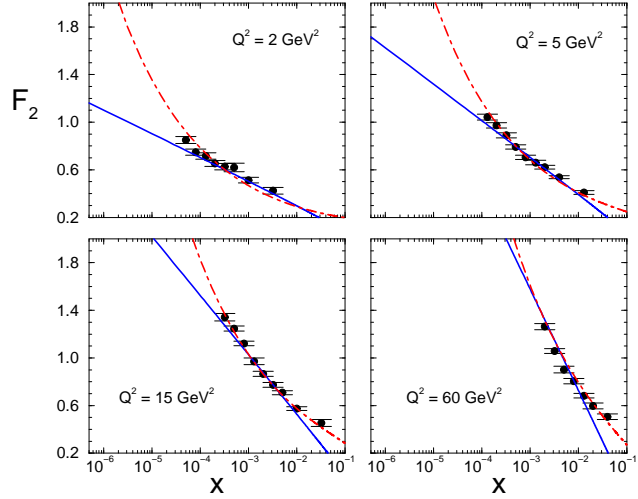,width=80mm}}
\caption{The prediction of the truncated BFKL series (solid lines) for the
proton structure function versus $x$ compared with the complete BFKL approach
\cite{navelet} (dot-dashed lines) extrapolated to the  THERA kinematical
domain.}  \label{extrap} \end{figure}

An important calculation is to estimate the proton structure function in
the THERA region. Due to the quite different behavior for the gluon
distribution predicted by the complete BFKL series and its truncation, we
expect to distinguish the relevance of each additional order in the
perturbative expansion using the kinematical domain to be realized at THERA.
In Fig. (\ref{extrap}), we present the structure function versus $x$ at
representative virtualities plotted with the latest H1 data for this observable
\cite{preldata}, extrapolated up to momentum fraction $x=10^{-7}$. As a
remark, we have corrected the $F_2$ normalization from the truncated BFKL
series multiplying it by a factor 1.2 to take into account the background
effects (about 20 \% in this domain).

An impressive  deviation from the complete BFKL series, which takes into
account the $n$-gluon emission cascade,  have been found at very low $x$.  The
growth of the structure function, and implicitly of  the gluon distribution, is
softer than the solution from  DGLAP evolution equation and of the BFKL-like
models. The difference is stronger at lower virtualities, whereas at
larger $Q^2$ both approaches are in better agreement. Such a result is already
expected, since the most important region for saturation (unitarization) lies
in the domain of small $x$ and low $Q^2$.  As a final comment, when the data
from THERA are to be available the dynamics should be 
unambiguously distinguished. As a consequence, the observables depending
directly from the gluon distribution, as $F_L$ and charm structure function,
are expected to give a signal for clear deviations from the conventional
renormalization group evolutions, i.e. the DGLAP approach.

\section{Conclusions}
We study in detail the contribution of a finite sum of gluons ladders to the
deep inelastic process, considering the truncation of the complete BFKL
Pomeron series at the second order in the perturbative expansion $\alpha_s\ln
(1/x)$. The main motivation is that in the present $ep$ acellerator energies
there is no phase space in rapidity $y=\ln(1/x)$ to enable a $n$-gluon ($n
\rightarrow \infty$) emission cascade in the final state. An important and
additional question is the violation of the  unitarity bound in asymptotic
energies from the current pQCD approaches, i.e. DGLAP and BFKL. The truncation
provides a cross section unitarity safe, meaning that we are in a
pre-asymptotic regime. One more term in the expansion may be added as the
energy increases, which then should be enough at very high energies since $\sigma_{tot}
\sim \ln^2(1/x)$, reaching the Froissart bound.

Using the truncated BFKL we are able to calculate the proton structure
function and its $Q^2$ logarithmic derivative. The expressions are obtained
with three adjustable  parameters. Two of them came from the non-perturbative
physics taking into account the modeling to the Pomeron-proton coupling; the
${\cal N}_p$ gives the normalization of the proton impact factor and the scale
$\mu^2$ is associated to the radius of the gluonic form factor of the proton.
The energy scale $x_0$ provides the threshold region where the perturbative
calculation is reliable, and a theoretical expectation is that it takes the
value $x_0 \leq 10^{-2}$ (setting the small Bjorken $x$ region and the domain
of validity of the approach).

We choose to determine the parameters from a smaller data set,
meaning the the latest HERA measurements on the $Q^2$ logarithmic
derivative reported by the H1 and ZEUS Collaboration. The values found are
consistent with the naive estimates, namely $\mu^2$ lies in the
non-perturbative domain ($\mu \approx \Lambda_{QCD}$) in both data sets and 
$x_0$ takes a value consistent with the Regge limit. The fitted expression to
the slope describes this observable with good agreement, producing effectively
the same linear behavior in $\ln Q^2$ considered by the H1 Collaboration
experimental analyzes. Considering the $x$ dependence of the slope, we obtain
a gluon distribution softer than those coming from the usual approaches,
$xG(x,Q^2) \sim x^{-\lambda}$, when one goes towards small $x$. The
sensitivity in the growth of the gluon distibution at smaller momentum
fraction taking the next order in the expansion is  an interesting issue,
however we postpone such  analyzes for a future study.

Having the adjustable parameters obtained from the slope data, a
parameter-free study of the proton structure function is performed. Agreement
is obtained for the range $0.3 \leq Q^2 \leq 350$ GeV$^2$, considering the
kinematical constraint  $x \leq 10^{-2}$. If such a cut is not considered, the
description is restricted into a smaller range of virtualities ($0.3 \leq Q^2
\leq 150$ GeV$^2$).  The growth of the structure function  presents large
deviations from the steep increasing present at both the LO BFKL series and
the DGLAP approach, where $F_2 \sim x^{-\lambda}$ as already discussed above.
The non-perturbative contribution (from the soft dynamics), mainly at low
$Q^2$ virtualities, is found to be  small. Indeed, it  was estimated that such
effects introduces a  correction of about 20 \% in the overall normalization.
We are not able to predict reliably the measurements at very large $Q^2$ ,
which lie often at $x > 10^{-2}$ and where the approach is expected not to
hold. Moreover, the complete BFKL series and its  truncation up to second
order can not be distinguished  unambiguously  since both approaches describe
the current HERA data. Nevertheless, our  analyzes allowed to extend the
estimates to very low virtualities  $Q^2 \sim 0.3$ GeV$^2$ and larger $Q^2
\geq 150$ GeV$^2$.

The forthcoming THERA experiment opens a new kinematical window towards
smaller $x$ ($\sim 10^{-6}$) and higher virtualities. The predictions from the
saturation models may be tested in this domain, determining if unitarity
effects start to be important in the description of the main observables. The
extrapolation of our results in the THERA region produced the following
conclusions: the approach does not violate strongly the unitarity limit for
the logarithmic slope. The region where this bound and the approach meet
defines the domain of validity  of the last one. Extrapolating the proton
structure function down to momentum fraction $x \geq 10^{-7}$, remarkable
deviations from the complete BFKL series have been found at very low $x$. The
difference is stronger at lower virtualities, which allows to distinguish the
relevance of each additional term in the expansion,  when the data from THERA
are to be avaluable.

\section*{Acknowledgments}

M.V.T.M. acknowledges enlightening discussions with V.P. Gon\c{c}alves
concerning high density QCD and its estimates for the THERA region.  This work
was partially financed by CNPq and by PRONEX (Programa de Apoio a N\'ucleos de
Excel\^encia), Brazil.

\end{document}